\documentclass{article}
\usepackage{ijcai05}
\usepackage{latexsym}
\usepackage{times}
\long\def\COMMENT#1\ENDCOMMENT{\message{(Commented text...)}\par}

\def\ignore#1{}

\def\beqa{\begin{eqnarray}}
\def\eeqa{\end{eqnarray}}

\def\Box{\hspace*{\fill}\framebox{\makebox[\depth]{\ }}}

\def\calS0{{\cal S}_0}

\def\cald{{\cal D}}
\def\calq{{\cal Q}}

\newtheorem{definition}{Definition}[section]
\newtheorem{theorem}{Theorem}[section]

\newtheorem{proposition}{Proposition}[section]
\newtheorem{lemma}{Lemma}[section]

\newtheorem{example}{Example}

\newenvironment{proof}{\noindent{\em Proof.}}{\hfill\Box}

\def\wnnw#1{\omega_{#1}}

\def\tc#1{TC({#1})}
\def\union#1#2{{#1}\cup{#2}}
\def\prior#1#2{{#1}\rhd{#2}}
\def\sdif#1#2{{#1}\ominus{#2}}

\title{{\bf Monotonic and Nonmonotonic Preference Revision}
\thanks{Research supported by NSF grant IIS-0307434.}
}
\author{Jan Chomicki \hspace{.5in} Joyce Song\\
University at Buffalo\\
{\tt $\{$chomicki,jsong5$\}$@cse.buffalo.edu}}

\begin{document}

\maketitle

\begin{abstract}
We study here preference revision, considering both the {\em monotonic} case where
the original preferences are preserved and the {\em nonmonotonic} case where the new preferences
may override the original ones.
We use a relational framework in which preferences are represented using binary relations 
(not necessarily finite).
We identify several classes of revisions that preserve order axioms, 
for example the axioms of strict partial or weak orders.
We consider applications of our results to preference querying in relational databases.
\end{abstract}

\section{Introduction}\label{s:intro}
The notion of {\em preference} is common in various contexts involving 
decision or choice. Classical utility theory \cite{Fish70} views preferences
as {\em binary relations}. A similar view has recently been
espoused in database research \cite{ChTODS03,Kie02,KiKo02}, where preference relations
are used in formulating {\em preference queries}. In AI, various approaches
to compact specification of preferences have been explored \cite{BouetalJAIR04}. The semantics
underlying such approaches typically relies on preference relations between worlds.

However, user preferences are rarely static \cite{PuFaTo03}. A database user may be disappointed
by the result of a preference query and decide to revise the preferences in the
query. In fact, a user may start with a partial or vague concept of her preferences,
and subsequently refine that concept.
An agent may learn  more about its task domain and consequently revise its preferences.
Thus, it is natural to study {\em preference revision}, as we do in the present paper. 

Preference revision shares some of the principles, namely  minimal change and primacy of new information,
with classical belief revision \cite{GaRo95}. However, its basic setting  is different.
In belief revision, propositional theories are revised with propositional formulas, yielding
new theories. In preference revision, binary preference relations are revised with other
preference relations, yielding new preference relations.
Preference relations are single, finitely representable (though possibly infinite) first-order structures, 
satisfying order axioms.

We distinguish between {\em monotonic} and {\em nonmonotonic} preference revision.
In the former, the original preference relation is fully incorporated into the revised
one. In the latter, the original preference relation may conflict with the revising
relation, leading to the necessity of retracting some of the original preferences.
We focus on two special cases: {\em refinement} in which both the original and the revising
relation are preserved, and {\em overriding revision} in which the revising relation may override
the original one. We adopt the notion of minimal change based on symmetric difference between
sets of tuples.

The challenges are: (1) to guarantee that suitable order properties,
for example the axioms of strict partial orders, are preserved by the
revisions, and (2) to obtain unique revisions.  Strict partial orders
(and weak orders), apart from being intuitive, enjoy a number of
attractive properties in the context of preference queries, as explained later in the
paper.  So it is desirable for revisions to
preserve such orders.  The uniqueness property is also important from
the user's point of view, as the user typically desires to obtain a
single revised preference relation. The presence of multiple revision
candidates necessitates some form of aggregation of or choice among
the candidates.  Fortunately, in the cases studied in this paper
there exist least revisions preserving the appropriate order axioms,
and thus uniqueness is obtained automatically.

We adopt the preference query framework of \cite{ChTODS03} (a similar model
was described in \cite{Kie02}), in which preference relations between tuples are defined by
logical formulas. \cite{ChTODS03} proposed a new relational algebra operator called {\em winnow} that
selects from its argument relation the {\em most preferred tuples} according
to the given preference relation.

\begin{example}\label{ex:car}
Consider the relation $Car(Make,Year)$
and the following preference relation $\succ_{C_1}$ between {\em Car} tuples:
\begin{quote}
{\em within each make, prefer a more recent car.}
\end{quote}
which can be defined as follows:
\[(m,y)\succ_{C_1}(m',y') \equiv m=m'\wedge y>y'.\]
The winnow operator $\wnnw{C_1}$ returns for every make the most recent car available.
Consider the instance $r_1$ of $Car$ in Figure \ref{fig:car}.
The set of tuples $\wnnw{C_1}(r_1)$ is shown in Figure \ref{fig:winnow}.

\begin{figure}[htb]
\centering
\begin{tabular}{|l|l|l|}
\hline
&{\em Make} &{\em Year} \\\hline
$t_1$& VW & 2002\\
$t_2$& VW& 1997 \\
$t_3$& Kia & 1997\\\hline
\end{tabular}
\caption{The Car relation}
\label{fig:car}
\end{figure}

\begin{figure}[htb]
\centering
\begin{tabular}{|l|l|l|}
\hline
&{\em Make} &{\em Year} \\\hline
$t_1$& VW & 2002\\
$t_3$& Kia & 1997\\\hline
\end{tabular}
\caption{The result of winnow}
\label{fig:winnow}
\end{figure}
\end{example}

\begin{example}\label{ex:car:1}
Example \ref{ex:car} provides a motivation for studying preference 
revision.
Seeing the result of the query $\wnnw{C_1}(r_1)$, a user may realize that the preference 
relation $\succ_{C_1}$ is not quite what she had in mind. The result of the query
may contain some unexpected or unwanted tuples, for example $t_3$.
Thus the preference relation needs to be modified, for example
by refining it with the following preference relation $\succ_{C_2}$:
\[(m,y)\succ_{C_2}(m',y') \equiv m={\rm ''VW''}\wedge m'\not={\rm ''VW''}\wedge y=y'.\]
The resulting refinement will contain both $\succ_{C_1}$ and $\succ_{C_2}$.
The tuple $t_3$ is now dominated by $t_2$ and will not be returned to the user.
\end{example}

In the terminology used in research on preference reasoning in AI \cite{BouetalJAIR04},
a relational database instance corresponds to the set of {\em feasible outcomes}
and the winnow operator picks the undominated (best) outcomes from this set, according
to the given preferences.
A preference setting can be affected by a change in preferences or a modification
of the set of possible outcomes. In this research, we address the former problem;
the latter one, database update, has been extensively studied in database research.
Moreover, we limit ourselves to preference revisions in which new preference 
information is combined, perhaps nonmonotonically, with the old one. We assume that the domains of preferences do not
change in revisions.

\ignore{
First, we introduce the basic notions of preference relations (based on \cite{ChTODS03}).
Then, we define various kinds of preference revision and discuss the preservation
of strict partial and weak orders by revisions.
Subsequently, we show that even when order axioms
are not guaranteed to be preserved, they can still be effectively checked.
Next, we show how the results of \cite{ChTODS03}
make it possible to incrementally compute preference query results
under iterated monotonic revisions.
Finally, we discuss related work and draw conclusions.
Some proofs are sketched. The remaining results can be proved by exhaustive case analysis.
}
\section{Basic notions}\label{sec:basic}
We are working in the context of the relational model of data.
Relation schemas consist of finite sets of attributes.
For concreteness, we consider two infinite domains: $\cald$ (uninterpreted constants, for readability shown as strings) and $\calq$ (rational numbers), but our results, except where explicitly indicated,
hold also for finite domains.
We assume that database instances are finite sets of tuples.
Additionally,
we have the standard built-in predicates.

\subsection{Preference relations}

We adopt here the framework of \cite{ChTODS03}.

\begin{definition}\label{def:prefrel}
Given a relation schema $R(A_1 \cdots A_k)$
such that $U_i$, $1\leq i\leq k$, is the domain (either $\cald$ or $\calq$)
of the attribute $A_i$, a relation $\succ$ is a {\em preference relation over $R$}
if it is a subset of $(U_1\times\cdots\times U_k)\times (U_1\times\cdots\times U_k)$.
\end{definition}

Although we assume that database instances are finite, in the presence of
infinite domains preference relations can be infinite.


Typical properties of a preference relation $\succ$ include:
\begin{itemize}
\item {\em irreflexivity}: $\forall x.\ x\not\succ x;$
\item {\em transitivity}: $\forall x,y,z.\ (x\succ y \wedge y\succ z)\Rightarrow x\succ z;$
\item {\em negative transitivity}: $\forall x,y,z.\ (x\not\succ y \wedge y\not\succ z)\Rightarrow x\not\succ z;$
\item {\em connectivity}: $\forall x,y.\ x\succ y\vee y\succ x \vee x=y;$
\item {\em strict partial order (SPO)} if $\succ$  is
irreflexive and transitive;
\item {\em weak order} if $\succ$ is a
negatively transitive SPO;
\item {\em total order} if $\succ$ is
a connected SPO.
\end{itemize}

\begin{definition}\label{def:prefformula}
A {\em preference formula (pf)} $C(t_1,t_2)$ is a first-order
formula defining a preference relation $\succ_C$ in the standard
sense, namely
\[t_1\succ_C t_2\;{\rm iff}\; C(t_1,t_2).\]
An {\em intrinsic preference formula (ipf)} is a preference formula that
uses only built-in predicates.
\end{definition}

By using the notation $\succ_C$ for a preference relation, we assume that there is
an underlying pf $C$.
Occasionally, we will limit our attention
to ipfs consisting of the following two kinds of atomic formulas
(assuming we have two kinds of variables: $\cald$-variables and $\calq$-variables):
\begin{itemize}
\item {\em equality  constraints}: $x=y$, $x\not=y$, $x=c$, or $x\not= c$, where
$x$ and $y$ are $\cald$-variables, and $c$ is an uninterpreted constant;
\item {\em rational-order constraints}: $x\theta y$ or $x\theta c$, where
\mbox{$\theta\in\{=,\not=,<,>,\leq,\geq\}$,} $x$ and $y$ are $\calq$-variables, and $c$ is a
rational number.
\end{itemize}
An ipf whose all atomic formulas are equality  (resp. rational-order)
constraints will be called an {\em equality} (resp. {\em rational-order}) ipf.
Clearly, ipfs are a special case of general constraints
\cite{CDB00}, and define {\em fixed}, although possibly infinite,
relations.

Every preference relation $\succ$ generates an indifference relation $\sim$:
two tuples $t_1$ and $t_2$ are {\em indifferent}
($t_1\sim t_2$) if
neither is preferred to the other one, i.e.,
$t_1\not\succ t_2$ and $t_2\not\succ t_1$.
We will denote by $\sim_C$ the indifference relation generated by $\succ_C$.

Composite preference relations are defined from simpler ones
using logical connectives. We focus on two basic ways of composing preference
relations: 
\begin{itemize}
\item {\em union}: 
\[t_1\ (\union{\succ_1}{\succ_2})\ t_2\ {\rm iff}\ t_1\succ_1 t_2\vee t_1\succ_2 t_2;\]
\item {\em prioritized composition} (where $\sim_1$ is the
indifference relation generated by $\succ_1$): 
\[t_1\ (\prior{\succ_1}{\succ_2})\ t_2\ {\rm iff}\  t_1\succ_1 t_2 \vee (t_1\sim_1 t_2\wedge t_1\succ_2 t_2).\]
\end{itemize}

We also consider transitive closure:
\begin{definition}\label{def:transitive}
The {\em transitive closure} of a  preference relation
$\succ$ over a relation schema $R$ is a  preference relation 
$TC(\succ)$ over $R$ defined as: 
\[(t_1,t_2)\in TC(\succ)\;{\rm iff}\; t_1\succ^n t_2\;
{\rm for\; some\;} n> 0,\]
where:
\[\begin{array}{l}
t_1\succ^1 t_2\equiv t_1\succ t_2\\
t_1\succ^{n+1}t_2\equiv\exists t_3.\ 
t_1\succ t_3\wedge t_3\succ^n t_2.\\
\end{array}\]
\end{definition}

\ignore{
Clearly, in general Definition \ref{def:transitive} leads to infinite formulas.
However, as shown in \cite{ChTODS03}, in the cases that we consider in this paper the preference
relation $\succ_{C*}$ will in fact be defined by a finite formula
(this is because transitive closure can be expressed as a terminating Datalog program).}

\subsection{Winnow}
We define now an algebraic operator that picks from a given relation the
set of the {\em most preferred tuples}, according to a given preference relation.
\begin{definition}\label{def:winnow}{\rm \cite{ChTODS03}}
If $R$ is a relation schema and $\succ$ a preference
relation over $R$,
then the {\em winnow operator} is written as $\wnnw{\succ}(R)$,
and for every instance $r$ of $R$:
\[\wnnw{\succ}(r)=\{t\in r\mid\neg \exists t'\in r.\ t'\succ t\}.\]
\end{definition}
If a preference relation is defined using a pf $C$, we write simply $\wnnw{C}$ instead of $\wnnw{\succ_C}$.
A {\em preference query} is a relational algebra query containing at least
one occurrence of the winnow operator.

\subsection{Preference revision}


The basic setting is as follows: We have a preference relation $\succ$ and revise it with a 
{\em revising\/} preference relation $\succ_0$ to obtain a {\em revised\/} preference relation $\succ'$.
We also call $\succ'$ a {\em revision\/} of $\succ$.
We limit ourselves to preference relations over the same schema.

The revisions are characterized by a number of different parameters:
\begin{itemize}
\item {\em axiom preservation}: what order axioms are preserved in $\succ'$;
\item {\em content preservation}: what preference relations are preserved in $\succ'$;
\item {\em ordering} (of revisions).
\end{itemize}

\begin{definition}
A revision $\succ'$ of $\;\succ$ with $\;\succ_0$ is:
\begin{itemize}
\item a {\em transitive (resp. SPO, a weak order)} revision if $\succ'$ is transitive
(resp. an SPO, a weak order);
\item a {\em monotonic\/} revision if $\ \succ\, \subseteq\, \succ'$;
\item a {\em refinement\/} revision ({\em refinement} for short) if \mbox{$\ \union{\succ}{\succ_0}\, \subseteq\, \succ'$};
\item an {\em overriding\/} revision if $\ \prior{\succ_0}{\succ}\, \subseteq\, \succ'$.
\end{itemize}
\end{definition}

A refinement is monotonic.
An overriding revision does not have to be monotonic because it may
fail to preserve $\succ$.

We order revisions using the symmetric difference ($\ominus$).

\begin{definition}\label{ref:closeness}
Assume $\succ_1$ and $\succ_2$ are two revisions of  a preference relation $\succ$ 
with a preference relation $\succ_0$. We say that $\succ_1$ is {\em closer} than $\succ_2$  to $\succ$
if $\sdif{\succ_1}{\succ}\, \subset\, \sdif{\succ_2}{\succ}$.
\end{definition}
\begin{definition}
A {\em minimal (resp. least)} revision of $\;\succ$ with $\succ_0$ is a revision 
that is minimal 
(resp. least) in the closeness
order among all revisions of $\;\succ$ with $\succ_0$.
\end{definition}

Similarly, we talk about least transitive refinements (or overriding revisions), 
least SPO (or  weak order) refinements or overriding revisions etc.
It is easy to see that if we consider only refinements or overriding revisions of a fixed
preference relation, closeness
reduces to set containment.

\begin{example}
Consider the  preference relation $\succ=\{(a,b),(b,c),(a,c)\}$
 representing the preference order $a\succ b\succ c$,
and the following revision of $\succ$, $\succ_1=\{(b,a),(b,c),(a,c)\}$. 
The revision $\succ_1$ is the least SPO overriding revision 
of $\,\succ$ with $\succ_0=\{(b,a)\}$. It achieves the effect of swapping $a$ and $b$
in the preference order.
\ignore{
Similarly, the revision $\succ_2$ is the least SPO overriding revision 
of $\,\succ$ with $\succ_0'=\{(c,a),(c,b),(b,a)\}$, achieving the effect of swapping $a$ and $c$.
In this paper we typically order by closeness the revisions of a fixed 
$\;\succ$ with a fixed $\;\succ_0$. However, closeness can also serve
to compare revisions of the same revised relation with {\em different} revising
relations. In the current example, we obtain 
that, as expected, $\succ_1$ is closer than $\succ_2$ to $\succ$,
because 
$\sdif{\succ_1}{\succ}\subset \sdif{\succ_2}{\succ}$.
}
\end{example}

To further describe the behavior of revisions, we define {\em preference conflicts}.
\begin{definition}
A {\em conflict} between a preference relation $\;\succ$  and a preference relation $\;\succ_0$ 
is a pair $(t_1,t_2)$ such that \mbox{$t_1\succ_0 t_2$} and \mbox{$t_2\succ t_1$}.
A {\em hidden conflict} between $\;\succ$  and $\;\succ_0$ is a pair $(t_1,t_2)$ such that $t_1\succ_0 t_2$ 
and there exist $s_1,\ldots s_k$, $k\geq 1$, such that
$t_2\succ s_1\succ\cdots\succ s_k\succ t_1$ and $t_1\not\succ_0 s_1 \not\succ_0\cdots\not\succ_0 s_k\not\succ t_2$.
\end{definition}
A hidden conflict is a conflict (if $\succ$ is an SPO) but not necessarily vice versa.
\begin{example}
If $\succ_0=\{(a,b)\}$ and $\succ=\{(b,a)\}$, then $(a,b)$ is a conflict which is not hidden. 
If we add $(b,c)$ and $(c,a)$ to $\succ$, then the conflict is also a hidden conflict ($s_1=c$).
If we further add $(c,b)$ or $(a,c)$ to $\succ_0$, then the conflict is not hidden anymore.
\end{example}

In this paper, we focus on refinement and overriding revisions because in our opinion
they capture two basic ways of revising preferences. A refinement does not retract
any preferences or resolve conflicts: it only adds new preferences necessitated by order properties.
So for a refinement to satisfy SPO properties, all conflicts need to be avoided.
An overriding revision, on the other hand, can override some of the original preferences
if they conflict with the new ones. Overriding can deal with conflicts which are not hidden
and solves all of them
in the same fashion: it gives higher priority to new preference information (i.e., $\succ_0$). 
Both refinement and overriding revisions preserve the revising relation $\succ_0$.

We now characterize those combinations of $\succ$  and $\succ_0$ that avoid all (or only hidden) conflicts.
\begin{definition}\label{def:compat}
A preference relation $\succ$  is {\em compatible} (resp. {\em semi-compatible}) with a preference relation
$\succ_0$  if there are no conflicts (resp. no hidden conflicts) between $\succ$  and $\succ_0$.
\end{definition}
Compatibility is symmetric and  implies semi-compatibility for SPOs. Semi-compatibility is not necessarily symmetric.
Examples \ref{ex:car} and \ref{ex:car:1} show a pair
of compatible relations. 
The compatibility of $\succ$ and $\succ_0$ {\em does not require} the acyclicity of $\succ\cup\succ_0$ or that one of the following hold:
$\succ\subseteq\succ_0$, $\succ_0\subseteq \succ$, or $\;\succ\cap\succ_0=\emptyset$.
For the former, consider $\succ=\{(a,b),(c,d)\}$ and $\succ_0=\{(b,c),(d,a)\}$.
For the latter, consider $\succ=\{(a,b),(b,c),(a,c)\}$ and $\succ_0=\{(a,b),(a,d)\}$.
\ignore{
A semi-compatible relation may conflict with a given preference relation. However,
in each such case, i.e., when $t_1\succ_0 t_2$ and $t_2\succ t_1$, all the ways of deriving  $t_2\succ t_1$
by transitivity have at least one pair of tuples in conflict with some pair of tuples in $\succ_0$.
}

All the properties listed above, including both variants of compatibility, are decidable for
equality or rational order ipfs.
For example, semi-compatibility is expressed by the condition
$\;\succ_0^{-1}\,\cap\, TC(\succ^{-1}\! -\! \succ_0^{-1})=\emptyset$ where $\succ^{-1}$ is the inverse
of the preference relation $\succ$.

\section{Preservation of order axioms}\label{sec:preserve}
We prove now a number of results that characterize refinement and overriding
revisions of 
of preference relations. The results are of the form:

\vspace{10pt}
\noindent
{\em Given that the original preference relation $\succ$ and the revising relation $\succ_0$ satisfy
certain order axioms, what kind of order axioms does the revision $\succ'$ satisfy?}

\vspace{10pt}
To capture minimal change of preferences, we typically study {\em least} revisions.
The revision setting helps to overcome the limitations of {\em preference composition}
\cite{ChTODS03} where it is shown that common classes of orders (SPOs, weak orders)
are often not closed w.r.t. basic preference composition operators
like union or prioritized composition. In the results that follow, we
obtain closure under least revisions thanks to (1) restricting $\succ$
and $\succ_0$, and (2) guaranteeing transitivity by
explicitly applying transitive closure where necessary.

\ignore{
(In the cases when preservation does not hold in general, it may still hold for specific
preference relations. Whether the latter is true or not can be effectively determined 
for preference relations that are definable using the constraint classes studied in this paper.)}

\ignore{
Our studies have shown that the transitive closure of the union of $\succ$ and $\succ_0$,
$TC(\succ\cup\succ_0)$, plays a special role as a refinement. It is the least transitive
refinement, so in the cases when it is irreflexive it becomes the least strict partial
order refinement. In some cases the union $\succ\cup\succ_0$ is already transitive, and thus
$TC(\succ\cup\succ_0)=\succ\cup\succ_0$.
}

\subsection{General properties}

\begin{lemma}\label{lem:equiv}
For compatible $\succ$ and $\succ_0$, 
\[\union{\succ_0}{\succ}=\prior{\succ_0}{\succ}.\]
\end{lemma} 

\begin{lemma}\label{lem:basic}
The preference relation $\union{\succ}{\succ_0}$ (resp.  $\prior{\succ_0}{\succ}$) is contained in every 
refinement (resp. overriding revision) of $\;\succ$ with $\succ_0$ and is, therefore, 
the least refinement (resp. least overriding revision) of $\;\succ$ with $\succ_0$.
\end{lemma}
\begin{lemma}\label{lem:TC}
The preference relation $\tc{\union{\succ}{\succ_0}}$ (resp. $\tc{\prior{\succ_0}{\succ}}$)
is contained in every transitive refinement (resp. every overriding revision) of $\;\succ$ with $\succ_0$ and is,
therefore, the least transitive refinement (resp. least transitive overriding revision) of $\;\succ$ with $\succ_0$.
\end{lemma}

\subsection{Strict partial orders}

SPOs have several important properties from the user's point of view, and thus
their preservation is desirable. For instance,
all the preference relations defined in \cite{Kie02} and the language Preference SQL \cite{KiKo02} are SPOs. 
Moreover, if $\succ$ is an SPO, then the winnow $\wnnw{\succ}(r)$ is nonempty 
if (a finite) $r$ is nonempty.
Also, the fundamental  algorithms for computing winnow require that the preference relation
be an SPO \cite{ChTODS03}.

In order to obtain the least SPO revisions, we have to make sure
that  $\tc{\union{\succ}{\succ_0}}$ and $\tc{\succ\rhd\succ_0}$ are irreflexive (they are transitive by definition). 

\ignore{
We need to impose a number of restrictions on $\succ$ and $\succ_0$.
First, we introduce the single-chain property. 
Second, we limit possible interactions between
those relations. Refinement requires the compatibility of $\succ$ and $\succ_0$, in order to 
eliminate the conflicts between $\succ$
and $\succ_0$.  Such conflicts, for example $a\succ b$ and $b\succ_0 a$,
immediately violate the irreflexivity of $TC(\succ\cup\succ_{0})$. 
For dealing with overriding revisions compatibility
can be replaced by a less restrictive condition, {\em semi-compatibility}, because prioritized composition
already provides a way of resolving some conflicts. In this way, nonmonotonicity of revisions becomes possible.
}

\begin{definition}
An SPO has the {\em single-chain property} (SCP) if it has at most
one maximal chain (maximal totally-ordered subset) having at least two elements.
Such a chain is called a {\em superchain}.
\end{definition}

The superchain in the above definition does not have to exhaust all the elements of
the domain, so an order having SCP does not have to be total or even weak. 

\begin{theorem}\label{th:spo:union}
For every compatible preference relations $\succ$ and $\succ_0$ such that
both are SPOs and at least one has SCP,
the preference relation $\, \tc{\union{\succ}{\succ_0}}$ 
is the least SPO refinement of $\;\succ$ with $\succ_0$.
\end{theorem}
\begin{proof} (sketch)
Assume $\succ$ has SCP.
If $\, \tc{\union{\succ}{\succ_0}}$ is not irreflexive, then $\union{\succ}{\succ_0}$
has a cycle. Consider such cycle of minimum length.
It consists of alternating $\succ_0$- and $\succ$-edges (otherwise it can be shortened).
If there is more than one $\succ$-edge in the cycle, then one of the
assumptions is violated. So the cycle consists of two edges:
$t_1\succ_0 t_2$ and $t_2\succ t_1$. But this is a conflict violating
compatibility.
\end{proof}

\begin{example}\label{ex:car:3}
Consider again the preference relation $\succ_{C_1}$:
\[(m,y)\succ_{C_1}(m',y') \equiv m=m'\wedge y>y'.\]
Suppose that the new preference information is captured as  $\succ_{C_3}$
which is a single-chain SPO:
\[\begin{array}{lcl}
(m,y)\succ_{C_3}(m',y')& \equiv &m={\rm ''VW''}\wedge y=1999\\
&\wedge &m'={\rm ''Kia''}\wedge y'=1999.\end{array}\]
Then $\tc{\union{\succ_{C_1}}{\succ_{C_3}}}$ is defined as the SPO $\succ_{C_4}$:
\[\begin{array}{l}
(m,y)\succ_{C_4}(m',y')\equiv m=m'\wedge y>y'\\\vee\
 m={\rm ''VW''}\wedge y\geq 1999\wedge m'={\rm ''Kia''}\wedge y'\leq 1999.
\end{array}\]
\end{example}

One can find examples where SCP or the compatibility
of $\succ$ and $\succ_0$ is violated, and a cycle
in $\union{\succ}{\succ_0}$ is obtained.

\ignore{
If the compatibility of $\succ$ and $\succ_0$ is relaxed, then it is easy
to obtain a cycle in $TC(\union{\succ}{\succ_0})$ and consequently, no strict partial
order refinement. SCP is also essential, as shown below.
\begin{theorem}\label{th:spo:greater}
There exist compatible preference relations $\succ$ and $\succ_0$ such that
(1) $\succ$ is an SPO and (2) $\succ_0$ is an SPO
of cardinality two, and 
there is no SPO refinement of $\succ$ with $\succ_0$.
\end{theorem}
\begin{proof}
Let a, b, c, and d tuples be in a database, $\succ=\{(a,b),(c,d)\}$ and
$\succ_{0}=\{(b,c),(d,a)\}$.
In this case, $TC(\union{\succ}{\succ_0})$  contains the tuple $(a,a)$ and thus is 
not irreflexive.
By Lemma \ref{lem:TC}, we know that  $TC(\union{\succ}{\succ_0})$ is 
contained in every transitive refinement. Therefore, every
transitive refinement is not irreflexive.
\end{proof}
}
\ignore{
Theorem \ref{th:spo:union} implies that if $\succ$ and $\succ_0$ are compatible and one of them
contains only one pair, then the least SPO refinement of $\succ$ with $\succ_0$ exists.
So what will happen if we break up the preference
relation $\succ_{0}$ from the proof of Theorem  
\ref{th:spo:greater} into two one-element relations $\succ_1$ and $\succ_2$ and attempt to 
apply Theorem \ref{th:spo:union} twice to get the least SPO
refinement?
Unfortunately, such a ``strategy'' does not work. The second
refinement is not possible because the preference relation $\succ_2$
is not compatible with the  refinement of $\succ$ with $\succ_1$.
}
For dealing with overriding revisions compatibility
can be replaced by a less restrictive condition, {\em semi-compatibility}, because prioritized composition
already provides a way of resolving some conflicts. 

\begin{theorem}\label{th:spo:priority}
For every preference relations $\succ$ and $\succ_0$ such that
both are SPOs, $\succ_0$ has SCP
and $\succ$ is semi-compatible with $\succ_0$,
the preference relation $\tc{\prior{\succ_0}{\succ}}$ 
is the least SPO overriding revision of $\;\succ$ with $\succ_0$.
\end{theorem}
\begin{proof} (sketch)
We assume that $\tc{\prior{\succ_0}{\succ}}$ is not irreflexive and  consider a cycle of minimum length 
in $\prior{\succ_0}{\succ}$.
This cycle has to consist of an edge $t_1\succ_0 t_2$ and a number of $\succ$-edges
$t_2\succ t_3,\ldots,t_{n-1}\succ t_n,t_n\succ t_1$ such that $n>2$.
(Here we cannot shorten sequences of consecutive $\succ$-edges because $\succ$
is not necessarily preserved in $\prior{\succ_0}{\succ}$.)
We have that $t_2\sim_0 t_3,\ldots,t_{n-1}\sim_0 t_n,t_n\sim_0 t_1$.
Thus $(t_1,t_2)$ is a hidden conflict violating the semi-compatibility of $\succ$
with $\succ_0$.
\end{proof}

Again, violating any of the conditions of Theorem \ref{th:spo:priority} may lead
to a situation in which no SPO overriding revision exists.
\ignore{
\begin{theorem}\label{th:spo:greater:priority}
There exist preference relations $\succ$ and $\succ_0$ such that
(1) $\succ$ is an SPO and (2) $\succ_0$ is an SPO,
and there is no SPO overriding revision of $\succ$ with $\succ_0$.
\end{theorem}
\begin{proof}
Consider a relation with three tuples $a$, $b$, and $c$.
Let $\succ_0=\{(b,a)\}$ and $\succ=\{(a,c),(c,b),(a,b)\}$.
Then there is a cycle in $\prior{\succ_0}{\succ}$.
\end{proof}
}
\begin{proposition}\label{prop:terminate}
For the preference relations defined using equality
or rational order ipfs,
the computation of \mbox{$TC(\union{\succ}{\succ_0})$} and \mbox{$TC(\succ\rhd\succ_0)$} terminates.
\end{proposition}
\ignore{
\begin{proof}
As pointed out in \cite{ChTODS03}, this follows from the termination
of an appropriate variant of Constraint Datalog \cite{CDB00}.
\end{proof}
}
The computation of transitive closure is done in a completely
database-independent way using Constraint Datalog techniques \cite{CDB00}. 
\begin{example}\label{ex:car:4}
Consider Examples \ref{ex:car} and  \ref{ex:car:3}. 
We can infer that
\[t_1=({\rm ''VW''},2002)\succ_{C_4} ({\rm ''Kia''},1997)=t_3,\]
because 
\[({\rm ''VW''},2002)\succ_{C_1}({\rm ''VW''},1999),\]
\[({\rm ''VW''},1999)\succ_{C_3}({\rm ''Kia''},1999),\]
and 
\[({\rm ''Kia''},1999)\succ_{C_1}({\rm ''Kia''},1997).\]
The tuples $({\rm ''VW''},1999)$ and $({\rm ''Kia''},1999)$ are {\em not\/} 
in the database.
\end{example}

\subsection{Weak orders}

\ignore{
We start by establishing a number of auxiliary results about weak orders.
By $\sim$ we denote the indifference relation generated by a preference relation $\succ$.
\begin{lemma}\label{lem:weak:1}
For every weak order preference relation $\succ$, and every x,y,
and z: if x$\sim$y and either x$\succ$z or y$\succ$z, then
x$\succ$z and y$\succ$z.
\end{lemma}
\begin{lemma}\label{lem:weak:2}
For every weak order preference relation $\succ$,
and every x,y,and
z: if x$\sim$y and either x$\not$$\succ$z or 
y$\not$$\succ$z, then x$\not$$\succ$z and y$\not$$\succ$z.
\end{lemma}
}
Weak partial orders are practically important because they capture the situation where
the domain can be decomposed into layers such that the layers are totally ordered and all the elements in one layer are
mutually indifferent. This is the case, for example, if the preference relation can be represented using a numeric
utility function.
If the preference relation is a weak order, a particularly efficient (essentially single pass)
algorithm for computing winnow is applicable \cite{ChCDB04}.

We first consider combinations of SPOs and weak orders.

\begin{theorem}\label{th:spo:weak}
For every compatible preference relations $\succ$ and $\succ_0$ such that
one is an SPO and the other
a weak order, the preference relation
$\union{\succ}{\succ_{0}}$ is the least SPO refinement of $\;\succ$ with $\succ_0$.
\end{theorem}
\ignore{
\begin{proof}
The relation $\union{\succ}{\succ_0}$ is irreflexive because  both
$\succ$ and $\succ_0$ are irreflexive.
Transitivity is proved by case analysis using Lemma \ref{lem:weak:1}.
\end{proof}
Clearly, the refinement yielding $\union{\succ}{\succ_0}$ under the assumptions of
Theorem \ref{th:spo:weak} cannot always be a weak order.
As a counterexample, consider for $\succ$ any SPO which is not a weak order,
for example:
\[a\succ b, a\sim c, b\sim c,\]
and for $\succ_0$ an empty order. 
Notice that in the above example the preference relation $\succ$ has  multiple minimal weak order 
extensions which are also minimal weak order refinements. Therefore, there
is no least weak order refinement.
}
In the context of overriding revisions, the requirement of compatibility becomes
unnecessary.
\begin{theorem}\label{th:spo:weak:priority}
For every preference relations $\succ_0$ and $\succ$ such that
$\succ_0$ is a weak order and $\succ$ an SPO, the preference relation
$\prior{\succ_0}{\succ}$ is the least SPO overriding revision of $\;\succ$ with $\succ_0$.
\end{theorem}

We consider now combinations of weak orders.

\begin{theorem}\label{th:weak:weak}
For every compatible weak order preference relations $\succ$ and $\succ_0$,
$\union{\succ}{\succ_{0}}$ is the least weak
order refinement of $\;\succ$ with $\succ_0$.
\end{theorem}
\ignore{
\begin{proof}
By Theorem \ref{th:spo:weak}, we know that the preference relation $\union{\succ}{\succ_0}$
is an SPO.
Negative transitivity of this relation is shown by
case analysis using Lemma \ref{lem:weak:2}.
\end{proof}
}
Again, for overriding revisions, we can relax the compatibility assumption.
This immediately follows from the fact that weak orders are closed with respect
to prioritized composition \cite{ChTODS03}.
\begin{proposition}\label{prop:weak:weak:priority}
For every weak order preference relations $\succ$ and $\succ_0$, the preference relation
$\prior{\succ_0}{\succ}$ is the least weak
order overriding revision of $\;\succ$ with $\;\succ_0$.
\end{proposition}

A basic notion in utility theory is that of {\em representability} of preference relations
using numeric utility  functions:

\begin{definition}\label{def:represent}
A real-valued function $u$ over a schema $R$ 
{\em represents} a preference relation $\succ$ over $R$ iff
\[\forall t_1,t_2\ [t_1\succ t_2 \;{\rm iff}\; u(t_1)>u(t_2)].\]
\end{definition}

Being a weak order is a necessary condition for the existence of a numeric representation
for a preference relation. However, it is not sufficient for uncountable orders \cite{Fish70}.
It is natural to ask whether the existence of  numeric representations
for the preference relations $\succ$ and $\succ_0$ implies the existence of such 
a representation for the least refinement $\succ'=(\union{\succ}{\succ_0})$.
This is indeed the case. 

\begin{theorem}\label{th:utility}
Assume that $\succ$ and $\succ_{0}$ are weak order preference
relations such that
\begin{enumerate}
\item $\succ$ and $\succ_{0}$ are compatible,
\item $\succ$  can be represented using a real-valued function $u$,
\item $\succ_{0}$ can be represented using a real-valued function $u_{0}$.
\end{enumerate}
Then $\succ'\, =\, \union{\succ}{\succ_{0}}$ is a weak
order preference relation that can be represented using any real-valued
function $u'$ such that for all $x$, $u'(x)=a\cdot u(x)+b\cdot u_0(x)+c$ where $a,b>0$.
\end{theorem}
\ignore{
\begin{proof}
By case analysis. The assumption of compatibility is essential.
\end{proof}
It is easy to see that $u'$ is not the only utility function representing $\succ'$.
Any positive linear combination of $u$ and $u_0$ has the same property.
}
Surprisingly, the compatibility requirement cannot in general be
replaced by semi-compatibility if we replace $\cup$ by $\rhd$ in Theorem \ref{th:utility}. This follows
from the fact that the lexicographic
composition of one-dimensional standard orders over ${\cal R}$ is
not representable using a utility function \cite{Fish70}.
Thus, preservation of {\em representability} is possible only 
under compatibility, in which case $\succ_0\rhd\succ\ = \ \succ_0\cup\succ$
(Lemma \ref{lem:equiv}) and the revision is monotonic.
It is an open question whether representability can be preserved under
nonmonotonic revisions.

We conclude this section by showing a general scenario in which the refinement of weak orders occurs in a natural way.
Assume that we have a numeric utility function $u$ representing a (weak order) preference relation $\succ$.
The indifference relation $\sim$ generated by $\succ$ is defined as:
\[x\sim y\ \equiv\ u(x)=u(y).\]
Suppose that the user discovers that $\sim$ is too coarse and needs to be further refined.
This may occur, for example, when $x$ and $y$ are tuples and the function $u$ takes
into account only some of their components. Another  function $u_0$ may be defined
to take into account other components of $x$ and $y$ (such components are called
{\em hidden attributes} \cite{PuFaTo03}). The revising preference relation $\succ_0$
is now:
\[x\succ_0 y\ \equiv\ u(x)=u(y)\wedge u_0(x)>u_0(y).\]
It is easy to see that $\succ_0$ is an SPO compatible with $\succ$ but not necessarily a weak order. 
Therefore, by Theorem \ref{th:spo:weak} the preference relation $\ \union{\succ}{\succ_0}\ $ is the least SPO
refinement of $\succ$ with $\succ_0$. 

\section{Checking axiom satisfaction}\label{sec:check}


If none of the results described so far implies that the least transitive
refinement of $\succ$ with $\succ_0$ is an SPO,
then this condition can often be explicitly checked.
Specifically, one has to: (1) compute the transitive closure $\tc{\union{\succ}{\succ_0}}$,
and (2) check whether the obtained relation is irreflexive.

From Proposition \ref{prop:terminate}, it follows that for equality and rational order ipfs
the computation of $\tc{\union{\succ}{\succ_0}}$ yields
some finite ipf $C(t_1,t_2)$. Then the second step reduces
to checking whether $\ C(t,t)$ is unsatisfiable, which is a decidable problem for equality and rational order
ipfs.

\begin{example}\label{ex:car:2}
Consider Examples \ref{ex:car} and \ref{ex:car:1}.
Neither of the preference relations $\succ_{C_1}$ and $\succ_{C_2}$ is a weak order or
has SCP.
Therefore, the results established earlier in this paper do not apply.
The preference relation $\succ_{C*}=\tc{\union{\succ_{C_1}}{\succ_{C_2}}}$ is defined
as follows:
\[\begin{array}{l}
(m,y)\succ_{C*}(m',y') \ \equiv\ m=m'\wedge y>y'\hspace{50pt}\\ 
\hfill\vee\ m={\rm ''VW''}\wedge m'\not={\rm ''VW''}\wedge y\geq y'\\ 

\end{array}\]
The preference relation $\succ_{C*}$ is irreflexive.
It also properly contains $\union{\succ_{C_1}}{\succ_{C_2}}$, because
$t_1\succ_{C*} t_3$ but $t_1\not \succ_{C_1}t_3$ and $t_1\not \succ_{C_2}t_3$.
The query $\wnnw{C*}(Car)$ evaluated in the instance $r_1$ (Figure \ref{fig:car}) returns only the tuple $t_1$. 
\end{example}

Similar considerations apply to overriding revisions and weak orders.
\ignore{
To determine whether a specific combination of $\succ$ and $\succ_0$
has the least weak order refinement (resp. least weak order overriding revision), 
one can check whether $TC(\union{\succ}{\succ_0})$ (resp. $TC(\prior{\succ_0}{\succ})$) is 
irreflexive and negatively transitive.
}
\ignore{
\section{Definability}
Theorems \ref{th:spo:weak}  and \ref{th:spo:weak:priority} demonstrates that 
if both $\succ$ and $\succ_0$ are definable, then
the least SPO refinement and overriding revision of $\succ$ with $\succ_0$ are definable
in the same language.
Similarly, Theorem \ref{th:weak:weak} and Proposition \ref{prop:weak:weak:priority} establish  definability 
in the weak order case.
However, Theorem \ref{th:spo:union} shows that in some cases the least SPO
refinement is not first-order definable, as its definition involves transitive closure.
It would be interesting to see if there are further cases in which definability can be achieved.
One of such cases is shown in Example \ref{ex:car:2}.
This problem is related to the problem of boundedness that was extensively studied
in the context of relational databases \cite{GaSaMaVa87}.
}
\section{Iterating monotonic preference revision}\label{sec:iterate}

Consider the scenario in which we iterate monotonic preference revision to obtain a sequence of preference relations 
$\succ_{1},\ldots,\succ_{n}$ such that each is an SPO and $\succ_{1}\,\subseteq\cdots\subseteq\, \succ_{n}$.
(Recall that refinement is monotonic but overriding revision not necessarily so.)
Assume that those relations are used to extract the best tuples from a fixed relation instance $r$.
Such evaluation provides feedback to the user about the quality of the given preference relation and may be
helpful in constructing its subsequent refinements.

In this scenario, the sequence of query results is:
\[r_0=r, r_1=\wnnw{\succ_1}(r),r_2=\wnnw{\succ_2}(r),\ldots,r_n=\wnnw{\succ_n}(r).\]

Proposition \ref{prop:contain} below implies that the sequence $r_0,r_1,\ldots,r_n$ is decreasing:
\[r_0\supseteq r_1\supseteq\cdots\supseteq r_n\]
and that it can be computed incrementally:
\[r_1=\wnnw{\succ_1}(r_0),r_2=\wnnw{\succ_2}(r_1),\ldots,r_n=\wnnw{\succ_n}(r_{n-1}).\]
To compute $r_i$, there is no need to look at the tuples in \mbox{$r-r_{i-1}$,} nor to
recompute winnow from scratch. The sets of tuples $r_1,\ldots,r_n$ are likely to 
have  much smaller cardinality than $r_0=r$.

\begin{proposition}{\rm \cite{ChTODS03}\label{prop:contain}}
If $\succ_{1}$ and $\succ_{2}$ are preference relations over a relation schema $R$ 
and \mbox{$\succ_{1}\, \subseteq\, \succ_{2}$}, 
then for all instances $r$ of $R$:
\begin{itemize} 
\item $\wnnw{\succ_2}(r)\,\subseteq\, \wnnw{\succ_1}(r);$
\item $\wnnw{\succ_2}(\wnnw{\succ_1}(r))\, =\, \wnnw{\succ_2}(r)$
if $\succ_1$ and $\succ_2$ are SPOs.
\end{itemize}
\end{proposition}

\ignore{
\section{Summary}
Tables \ref{tab:refinement} and \ref{tab:overriding} summarize the closure properties under refinement
and overiding revisions of different classes of preference relations.
\begin{table}[htb]
\centering
\begin{tabular}{|c|c|c|}
\hline
& $\succ$ SPO  & $\succ$ weak order \\\hline
$\succ_0$ SPO      & compatible: SPO if $\succ$ or $\succ_0$ single-chain  & compatible: SPO\\\hline
$\succ_0$ weak order      & compatible: SPO &compatible: weak order\\\hline
\end{tabular}
\caption{Closure under refinement}
\label{tab:refinement}
\end{table}
\begin{table}[htb]
\centering
\begin{tabular}{|c|c|c|}
\hline
& $\succ$ SPO  & $\succ$ weak order \\\hline
$\succ_0$ SPO      & semi-compatible: SPO if $\succ_0$ single-chain  & compatible: SPO\\\hline
$\succ_0$ weak order      & SPO &weak order\\\hline
\end{tabular}
\caption{Closure under overriding revision}
\label{tab:overriding}
\end{table}
}

\section{Related work}\label{sec:related}
CP-nets \cite{BouetalJAIR04} are an influential recent formalism for reasoning with conditional
preference statements under {\em ceteris paribus} semantics (such semantics is also adopted in other work
\cite{McGDo04,WeDo91}).
We conjecture that CP-nets can be expressed in the framework of preference relations of \cite{ChTODS03}, used
in the present paper, by making the semantics explicit.
If the conjecture is true, the results of the present paper will be relevant to revision
of CP-nets.
\begin{example}
The CP-net \mbox{$M=\{a\succ \bar{a}, a:b\succ \bar{b}, \bar{a}: \bar{b}\succ b\}$}
where $a$ and $b$ are Boolean variables,
captures the following preferences:
(1) prefer $a$ to $\bar{a}$, all else being equal;
(2) if $a$, prefer $b$ to $\bar{b}$; 
(3) if $\bar{a}$, prefer $\bar{b}$ to $b$.
We construct a preference relation $\succ_{C_M}$ between worlds, i.e., Boolean valuations of $a$ and $b$:
\[
\begin{array}{lcl}
(a,b)\succ_{C_M} (a',b') &\equiv &a=1 \wedge a'=0 \wedge b=b'\\&\vee & a=1 \wedge 
a'=1 \wedge b=1 \wedge b'=0 \\&\vee & a=0\wedge a'=0 \wedge b=0 \wedge b'=1.
\end{array}\]
Finally, the semantics of the CP-net is fully captured as the transitive closure $\;\tc{\succ_{C_M}}$.
Such closure can be computed using Constraint Datalog with Boolean constraints \cite{CDB00}.
\end{example}
CP-nets and related formalisms cannot express preference relations over infinite domains
which are essential in database applications.

\cite{PuFaTo03} formulates different scenarios of preference revision and does not contain any formal framework.
\cite{Freu04} describes minimal change revision of {\em rational} preference relations between propositional formulas.
We are not aware of any work on revising infinite preference relations.

\section {Conclusions and future work}\label{sec:concl}

We have presented a general framework for revising preference relations
and established a number of order axiom preservation results for specific classes of revisions.
In the future, we plan to consider more general classes of revisions
and databases with restricted domains, e.g., Boolean.
Another direction is the design of a {\em revision language}
in which different parameters of preference revision can be explicitly specified 
by the user.
Connections to {\em iterated belief revision} \cite{DaPe97} should also
be explored.
\ignore{
For more general classes of revisions than those studied in this paper,
the interplay between different requirements imposed on
revisions will be more complicated.  For example, in the presence of
multiple minimal revisions, the issues of
aggregating minimal revisions or choosing one among them will become
important. It will be also interesting to see whether considering specific
domains, for example finite domains, would make it possible to establish
new closure results for various classes of revisions.
}
\begin{small}

\end{small}

\end{document}